\documentstyle[twocolumn,epsf,prl,aps]{revtex}

\def\l{\langle}
\def\r{\rangle}

\begin{document}
\draft
\title{
Cluster Analysis and Finite-Size Scaling for Ising Spin Systems
}

\author{Yusuke Tomita and Yutaka Okabe\cite{okabe}}
\address{
Department of Physics, Tokyo Metropolitan University,
Hachioji, Tokyo 192-0397, Japan
}

\author{Chin-Kun Hu\cite{huck}}
\address{
Institute of Physics, Academia Sinica, Nankang, Taipei 11529, Taiwan
}

\maketitle

\begin{abstract}
Based on the connection between the Ising model and a correlated
percolation model, 
we calculate the distribution
function for the fraction ($c$) of lattice sites in percolating 
clusters in subgraphs with $n$ percolating clusters, $f_n(c)$, 
and the distribution function for magnetization ($m$) in subgraphs 
with $n$ percolating clusters, $p_n(m)$. We find that  
$f_n(c)$ and $p_n(m)$ have very good finite-size 
scaling behavior and they have universal finite-size scaling functions 
for the model on square, plane triangular, and honeycomb lattices
when aspect ratios of these lattices have the proportions 
1:$\sqrt 3$/2:$\sqrt 3$.
The complex structure of the magnetization distribution function 
$p(m)$ for the system with large aspect ratio could be understood 
from the independent orientations
of two or more percolation clusters in such system.  
\end{abstract}


\narrowtext

Universality and scaling are two important concepts in the modern era
of critical phenomena \cite{stanley71} and for analyzing the simulation
or experimental data of finite critical systems, one often appeals to
finite-size scaling \cite{fisher70,pf84,sa94} where both critical
exponents and finite-size scaling function play important role.
The universality of critical exponents was well known for a long time 
\cite{stanley71},
but the universality of finite-size scaling functions received much 
attention only in recent years
\cite{hlc95a,ok96,wh97,lhc98,hu98,ok99}. 
In 1984, Privman and Fisher first
proposed the idea of universal finite-size scaling functions (UFSSF's) with    
nonuniversal metric factors \cite{pf84}.
In 1995$\sim$1996, Hu, Lin and Chen (HLC) \cite{hlc95a} applied a
histogram Monte Carlo simulation method
\cite{hu92b} to calculate the existence probability \cite{hu92b}
(also called crossing probability \cite{lpps92}) $E_p$, the percolation
probability $P$, the probability for the appearance of $n$ percolating
clusters $W_n$ \cite{hlc95a} of site and bond
percolation on finite square (sq), plane triangular (pt), and honeycomb (hc)
lattices, whose aspect ratios approximately have the relative proportions
$1:\sqrt{3}/2:\sqrt{3}$ considered by Langlands, {\it et al.} \cite{lpps92}.
Using nonuniversal metric factors, HLC found that the six percolation 
models have very nice UFSSF's for $E_p$, $P$, and $W_n$ near the critical 
points \cite{hlc95a}, and at the critical point
the average number of percolating clusters increases linearly
with aspect ratios of the lattices \cite{hlc95a}.
Using Monte Carlo simulation,
Okabe and Kikuchi found UFSSF's for the Binder parameter $g$ \cite{Binder81}
and magnetization distribution functions $p(m)$
of the Ising model on planar lattices \cite{ok96},
and Wang and Hu found UFSSF's for dynamic critical phenomena
of the Ising model \cite{wh97}.  Based on the connection between 
the $q$-state bond correlated percolation model (BCPM)
and the $q$-state Potts model \cite{hu84} 
Hu, Chen, and Lin found UFSSF's for $E_p$ and $W_n$ of the $q$-state
BCPM without using nonuniversal metric factors \cite{hu98}.
It is of interest to get a deeper understanding
of the UFSSF's for $W_n$ and $P$ for the system with multiple
percolating clusters. 

In this paper, based on the connection between
the Ising model, i.e. the two-state Potts model, and 
the two-state BCPM, we use Monte Carlo method 
to calculate the distribution function
for the fraction ($c$) of lattice sites in percolating
clusters in subgraphs with $n$ percolating clusters, $f_n(c)$,
and the distribution function for magnetization
($m$) in subgraphs with $n$ percolating clusters, $p_n(m)$.
We find that  $f_n(c)$ and $p_n(m)$ have very good finite-size
scaling behavior and they have UFSSF's
for the model on sq, pt, hc lattices
when aspect ratios of these lattices have the proportions 
1:$\sqrt 3$/2:$\sqrt 3$.
Since $W_n$ and $P$ of the two-state BCPM for the Ising model 
may be calculated from $f_n(c)$, the universality of
finite-size scaling functions for $W_n$ and $P$ are related to 
universality of finite-size scaling functions for $f_n(c)$.  
The complex structure of $p(m)$ for the system
with large aspect ratio could be understood from the independent 
orientations of two or more percolation clusters in such system.
Our work suggests many problems for further research.

The Hamiltonian of the Ising model on an  $L_1 \times L_2$ lattice 
$G$ of $N_b$ bonds is given by
$  {\cal H} = -J \sum _{<i,j>} \sigma_i \sigma_j -h\sum_i \sigma_i$, 
where $\sigma_i=\pm 1$, $J > 0$ and is the ferromagnetic
coupling constant between the nearest-neighbor Ising spins, 
and $h$ is the external magnetic field. 
Using subgraph expansion,
Hu \cite{hu84} showed that the partition function of the Ising
model on $G$ may be written as
\begin{eqnarray}
 Z_N  &=& e^{KN_b}\sum_{G'\subseteq G}p^{b(G')}(1-p)^{N_b-b(G')} \cr
      &~& \quad \times \prod_{{\rm cluster}}[2\cosh(Bn_c(G'))] ,  
\label{Zn2}
\end{eqnarray}
where $p=1-e^{-2K}$, $K=J/(k_BT)$, $B=h/(k_BT)$, 
$b(G')$ is the number of occupied bonds in $G'$, and
the sum is over all subgraphs $G'$ of $G$, the product extends over all
clusters in a given $G'$, $n_c(G')$ is the number of sites in each cluster.
When $B=0$, Eq.~(\ref{Zn2}) reduces to the result of Ref. \cite{kf69}.  
The sites connected by occupied bonds are in the same cluster; 
all spins in a cluster must be in the same direction, which may be
up or down.
Using $Z_N$ of Eq.~(\ref{Zn2}), Hu found that the spontaneous magnetization
and the magnetic susceptibility of the Ising model are related to the
percolation probability $P$ and the mean cluster size of the 
(two-state) BCPM such that the probability weight for
the appearance of a subgraph $G'$ of $b(G')$ bonds and $n(G')$ clusters
is given by
$ \pi(G',p)= p^{b(G')}(1-p)^{N_b-b(G')}2^{n(G')}.$ 
Such connection ensures that the phase transition of the Ising model is
the percolation transition of the BCPM.
The extension of the probability weight $\pi(G',p)$ 
to a $q$-state Potts model is simply achieved 
by replacing 2 by $q$ in $\pi(G',p)$.
From $\pi(G',p)$ we can define 
\begin{equation}
 E_p(G,p)={\sum_{G'_p \subseteq G} \pi(G'_p,p) /
 \sum_{G' \subseteq G} \pi(G',p) },
\end{equation}
which is called the existence probability of the BCPM.
Here the sum in the denominator is over all subgraphs $G'$ of $G$
and the sum in the numerator is restricted to all percolating
subgraphs $G_p'$ of $G$.
In \cite{hu92c}, Hu and Chen found that $E_p(G,p)$ has
very good finite-size scaling behavior. 
In \cite{hu98}, Hu, Chen, and Lin found that $E_p(G,p)$
has UFSSF.

In the present paper we use the Wolff algorithm \cite{wolff88} for spin
update and  study the percolating properties of clusters on planar
lattices with periodic boundary conditions for both directions.  For the 
assignment of a bond-percolating cluster, we consider free and periodic 
boundary conditions in the vertical and horizontal directions, 
respectively; that is, a cluster which extends from the top row 
to the bottom row is a percolating cluster.

We first consider the fraction of lattice sites 
in the percolating clusters, $c$, and denote 
the probability distribution function of $c$ by $f(c)$. 
The average value of $c$ gives the percolation probability $P$, 
\begin{equation}
 \l c \r = \int_0^1 c f(c) dc = P ,
\label{eq_c}
\end{equation}  
and plays a role of order parameter in the percolation problem.
To study $c$ in subgraphs with exactly $n$ percolating clusters,
we decompose $f(c)$ by the number of percolating clusters,
$n$; that is, 
\begin{equation}
  f(c) = \sum_{n=1}^{\infty} f_n(c) .  
\end{equation}
We should note that 
\begin{equation}
  \int_0^1 f_n(c) dc = W_n  , \quad (n=1, \cdots, \infty)
\end{equation}
and 
\begin{equation}
  \sum_{n=1}^{\infty}W_n = 1-W_0 = E_p(G,p) .
\end{equation}
We may consider the quantity,
\begin{equation}
  \l c \r_n=\int_0^1 cf_n(c) dc , \quad (n=1, \cdots, \infty) , 
\label{eq_cn}
\end{equation}
which is the fraction of lattice sites in the $n$ percolating 
clusters.  In other words, we decompose $\l c \r$ by the 
number of percolating clusters such that 
\begin{equation}
  \l c \r = \sum_{n=1}^{\infty}\l c \r_n=P . 
\end{equation}

The probability distribution of the magnetization, $p(m)$, is 
an important quantity in the phase transition problem \cite{ok96,ok99}.  
Let us decompose $p(m)$ by the number of 
percolating clusters in the same way as in $f(c)$, 
\begin{equation}
  p(m) = \sum_{n=0}^{\infty}p_n(m) .  
\end{equation}
Now we have the relation, 
\begin{equation}
  \int_{-1}^1 p_n(m)dm = W_n  , \quad (n=0, \cdots, \infty) .  
\end{equation}

It should be noted that the relation between $f_n(c)$ and $p_n(m)$
is not a simple one especially for a system with multiple
percolating clusters.
There are two type of clusters, that is, the cluster
with up spins and that with down spins.  We may divide
the fraction of lattice sites in percolating clusters, $c$,
into two classes, $c_+$ and $c_-$.  By definition, $c = c_+ + c_-$
and to the leading order, $m \sim c_+ - c_-$.  
If there is only a single percolating cluster,
$m \sim c_+$ or $m \sim c_-$; thus, $m^2 \sim c^2$ in the leading
contribution.  However, if there are two or more percolating clusters,
the relation is not
trivial, and this is the origin of the complex structure
of $p(m)$ for the lattices with large aspect ratio $a$.
Therefore, the study of $c$ and $m$ becomes more interesting
in the case of multiple percolating clusters.  

According to the theory of finite-size scaling \cite{fisher70},  
if a quantity $Q$ has 
a singularity of the form $Q(t) \sim t^{\omega}$ near the criticality
$t=0$, then the corresponding quantity $Q(L,t)$ for 
the finite system with the linear size $L$ has a scaling form
$Q(L,t) \sim L^{-\omega/\nu} X(tL^{1/\nu})$, 
where $\nu$ is the correlation-length exponent and is 1 
for two-dimensional (2D) Ising model.
The finite-size scaling is also applicable to the distribution 
function of $Q$.  At the criticality $t=0$, we have 
a finite-size scaling form
$  p(Q;L,t=0) \sim L^{\omega/\nu} Y(Q L^{\omega/\nu})$. 
Thus,  we expect the following finite-size scaling relations:
$W_n(t) \sim X_n^a(tL^{1/\nu})$, 
$\l c \r_n(t) \sim L^{-\beta/\nu} X_n^b(tL^{1/\nu})$, 
$f_n(c;t=0) \sim L^{\beta/\nu} Y_n^a(c L^{\beta/\nu})$, 
$p_n(m;t=0) \sim L^{\beta/\nu} Y_n^b(m L^{\beta/\nu})$, 
where $\beta$ is the order-parameter exponent and is 1/8 for
2D Ising model.

The finite-size scaling functions usually
depend on the lattice, or other details of the system.
However, with appropriate choices of nonuniversal metric factors 
$D_1$ and $D_2$, 
\begin{eqnarray}
  D_2 Q(L,t) &=& L^{-\omega/\nu} \hat X(D_1 tL^{1/\nu}) , \\
  p(Q;L,t=0) &=& D_2 L^{\omega/\nu} \hat Y(D_2 Q L^{\omega/\nu}) , 
\end{eqnarray}
the finite-size scaling functions $\hat X, \hat Y$ could become universal.
This concept of the UFSSF was 
first proposed by Privman and Fisher \cite{pf84}, and 
has been recently confirmed for the percolation problem 
\cite{hlc95a,lhc98} and for the Ising model 
\cite{ok96,hu98}. We should note that the UFSSF's  
still depend on boundary conditions \cite{hlc95a}.

To study the finite-size scaling and the universality of $f_n(c)$, $f(c)$, 
$p_n(m)$, and $p(m)$, we calculate $f_n(c)$, $f(c)$, $p_n(m)$, $p(m)$, 
$W_n$, $\l c \r_n$, $\l c \r$,
$g_{cn}$, and $g_c$ for the BCPM on sq, pt, and hc lattices 
whose aspect ratios approximately have the proportions
1:$\sqrt 3/2$:$\sqrt 3$ and each kind of lattices has two linear
dimensions; here 
$  g_{cn} = {\l c \r_n}^2/{\l c^2 \r_n}$,
$  g_{c} = {\l c \r}^2/{\l c^2 \r}$,
and the second moments of $c$ are defined as in Eqs.~(\ref{eq_cn}) 
and (\ref{eq_c}).
We note that $g_{cn}$ and $g_c$ have the same finite-size scaling property
as the Binder parameter \cite{Binder81}.

\vspace{6mm}
\begin{figure}
\epsfxsize=\linewidth 
\centerline{\epsfbox{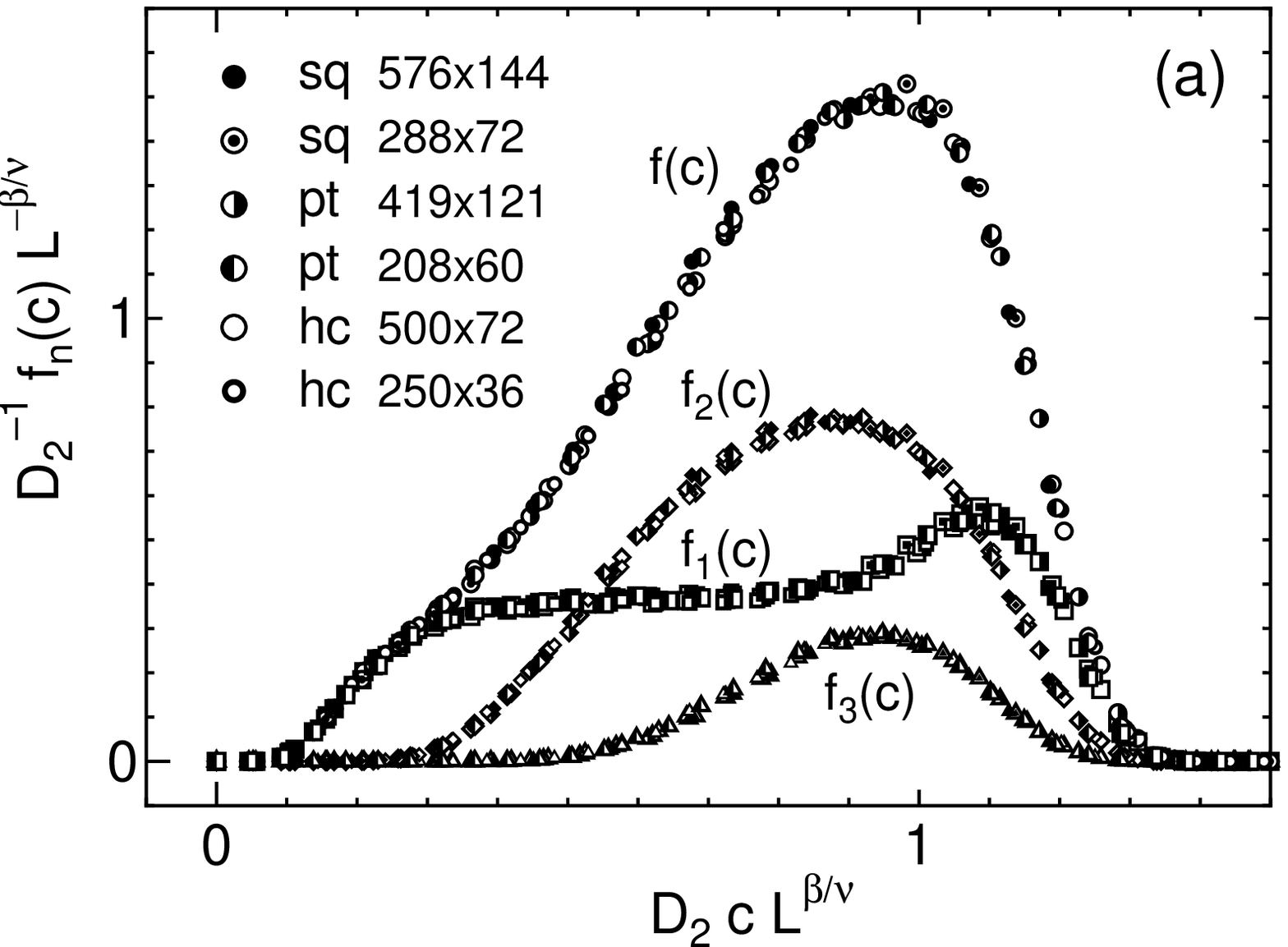}}
\vspace{6mm}
\epsfxsize=\linewidth 
\centerline{\epsfbox{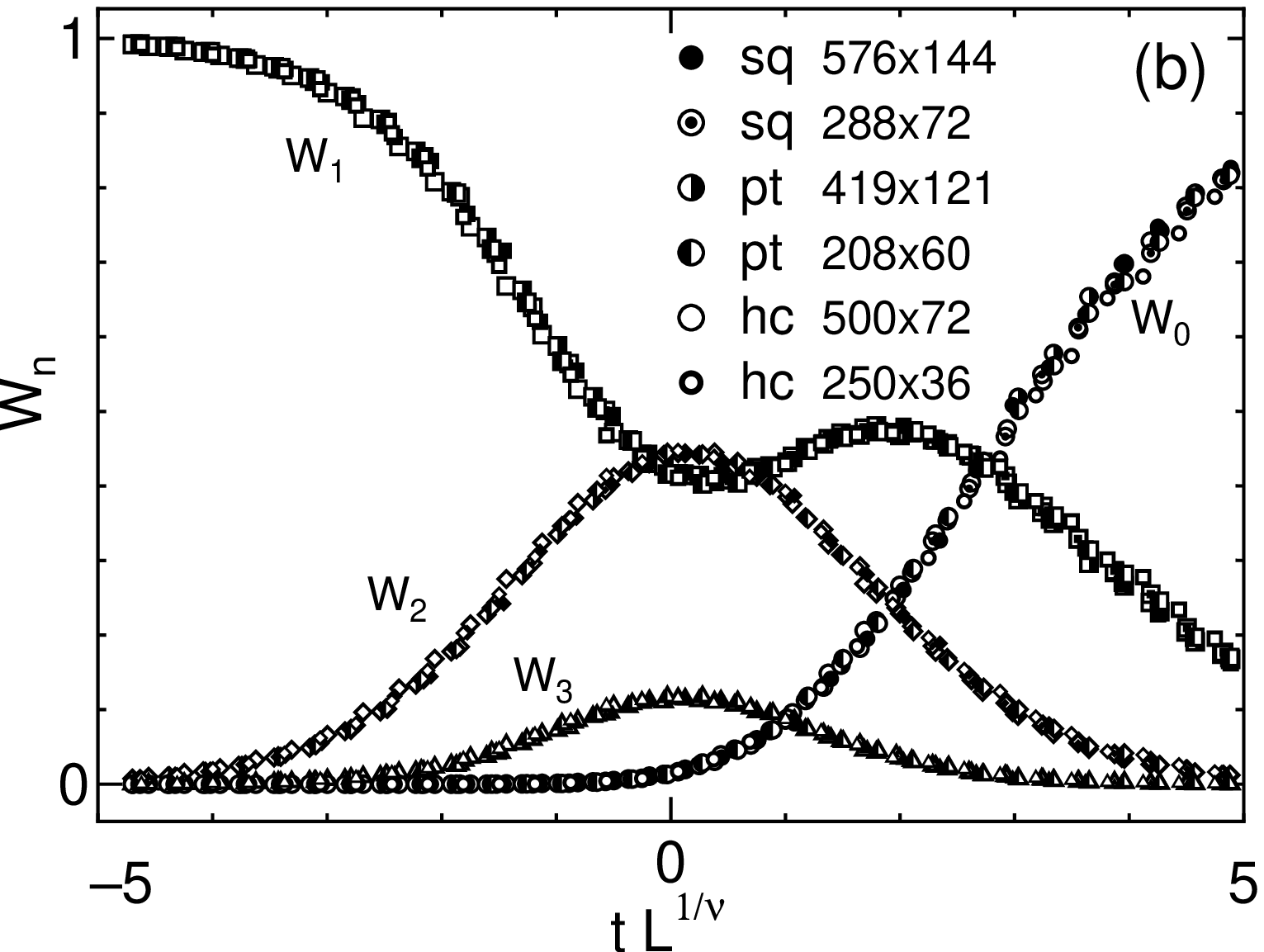}}
\vspace{6mm}
\epsfxsize=\linewidth 
\centerline{\epsfbox{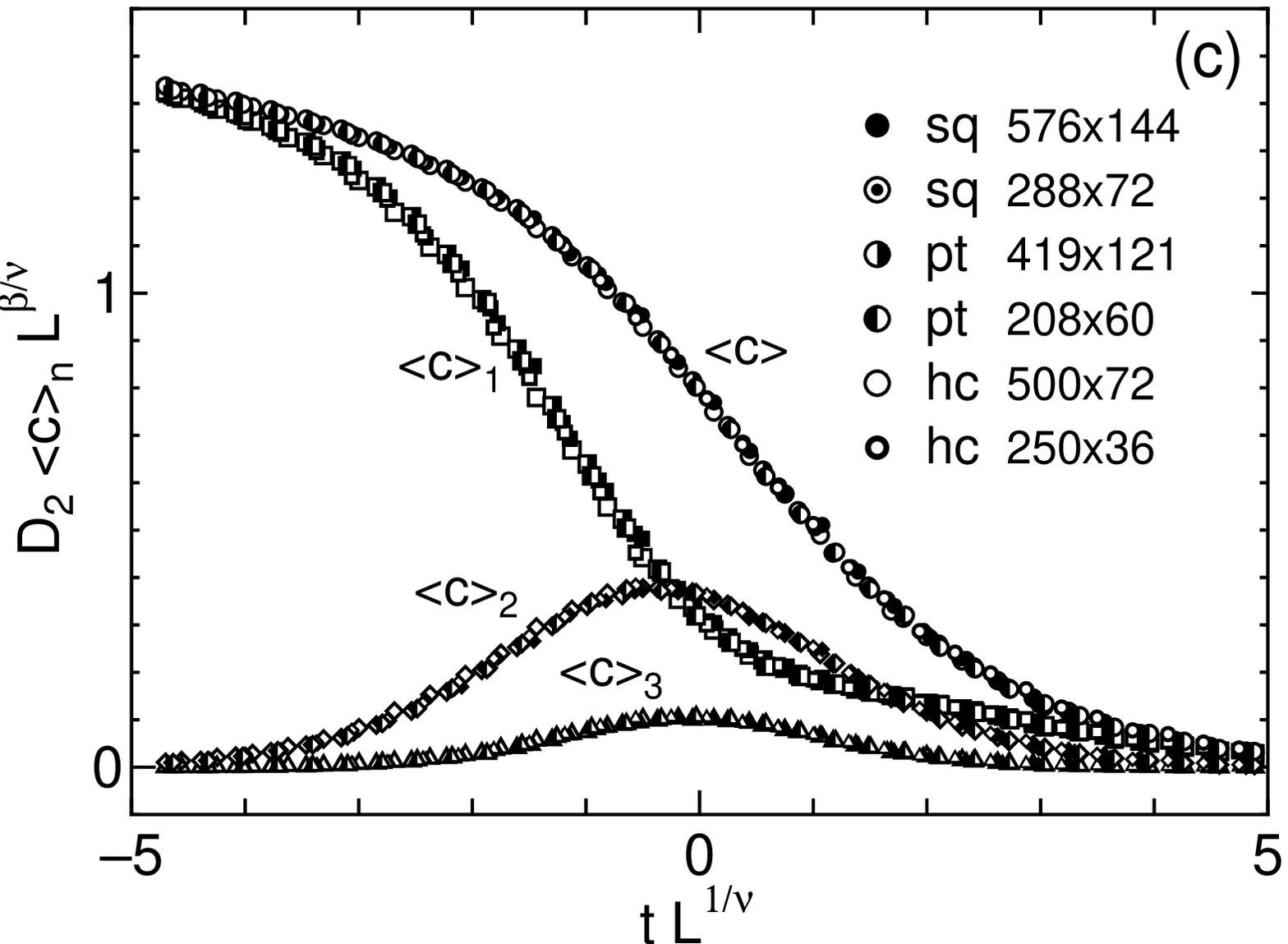}}
\vspace{6mm}
\epsfxsize=\linewidth 
\centerline{\epsfbox{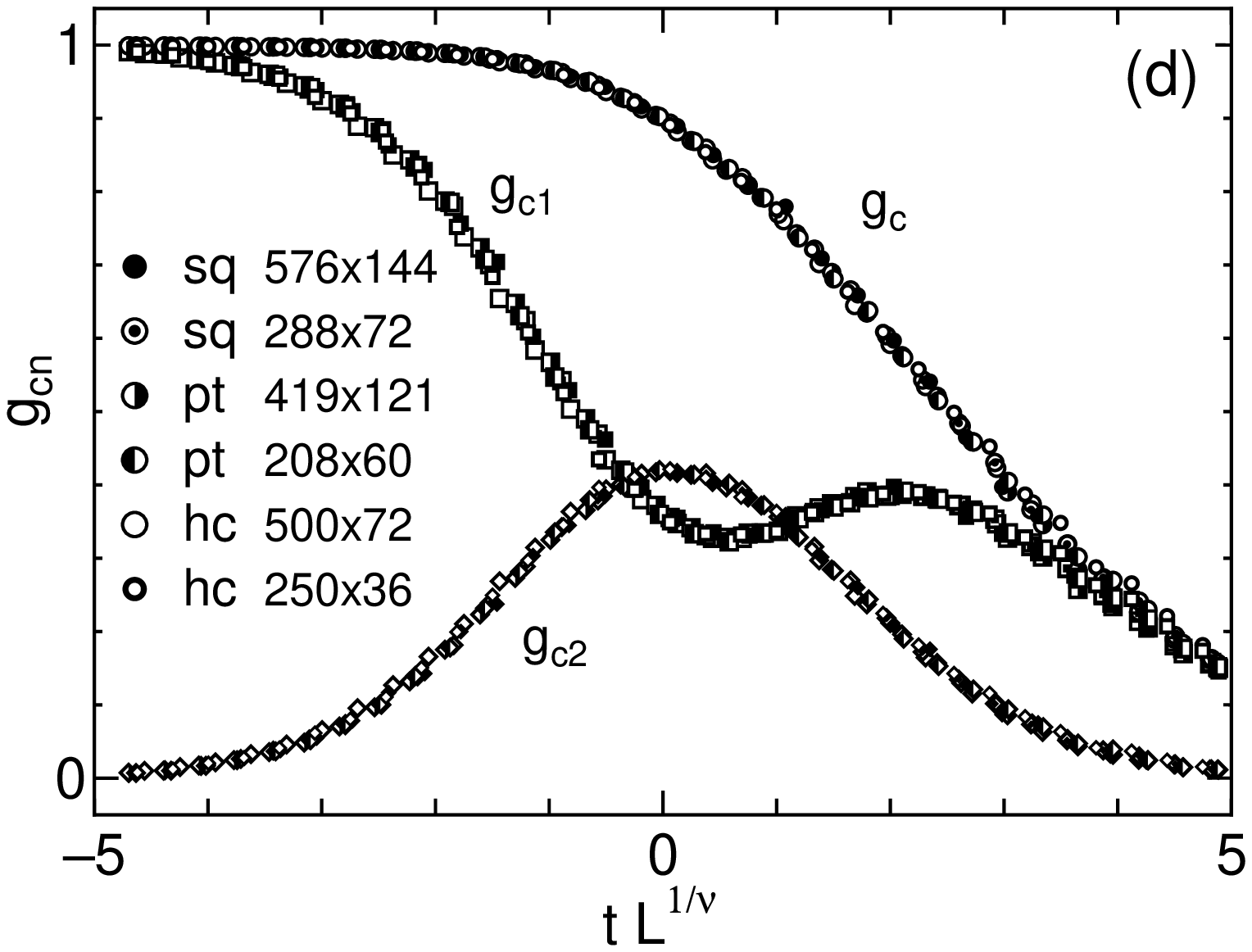}}
\vspace{2mm}
\caption{
Universal finite-size scaling functions for the Ising clusters
on the sq, pt, and hc lattices.
(a) $D_2^{-1} f_n(c)L^{-\beta/\nu}$ and $D_2^{-1} f(c)L^{-\beta/\nu}$ 
at the critical point as a function of $D_2 c L^{\beta/\nu}$,
(b) $W_n$ as a function of $tL^{1/\nu}$,
(c) $D_2 \l c \r_n  L^{\beta/\nu}$ and $D_2 \l c \r L^{\beta/\nu}$
 as a function of $tL^{1/\nu}$,
(d) $g_{cn} (=\l c \r_n^2/\l c^2 \r_n)$ and 
$g_c (=\l c \r^2/ \l c^2 \r)$ as a function of $tL^{1/\nu}$.
}
\label{fig1}
\end{figure}
\begin{figure}
\epsfxsize=\linewidth 
\centerline{\epsfbox{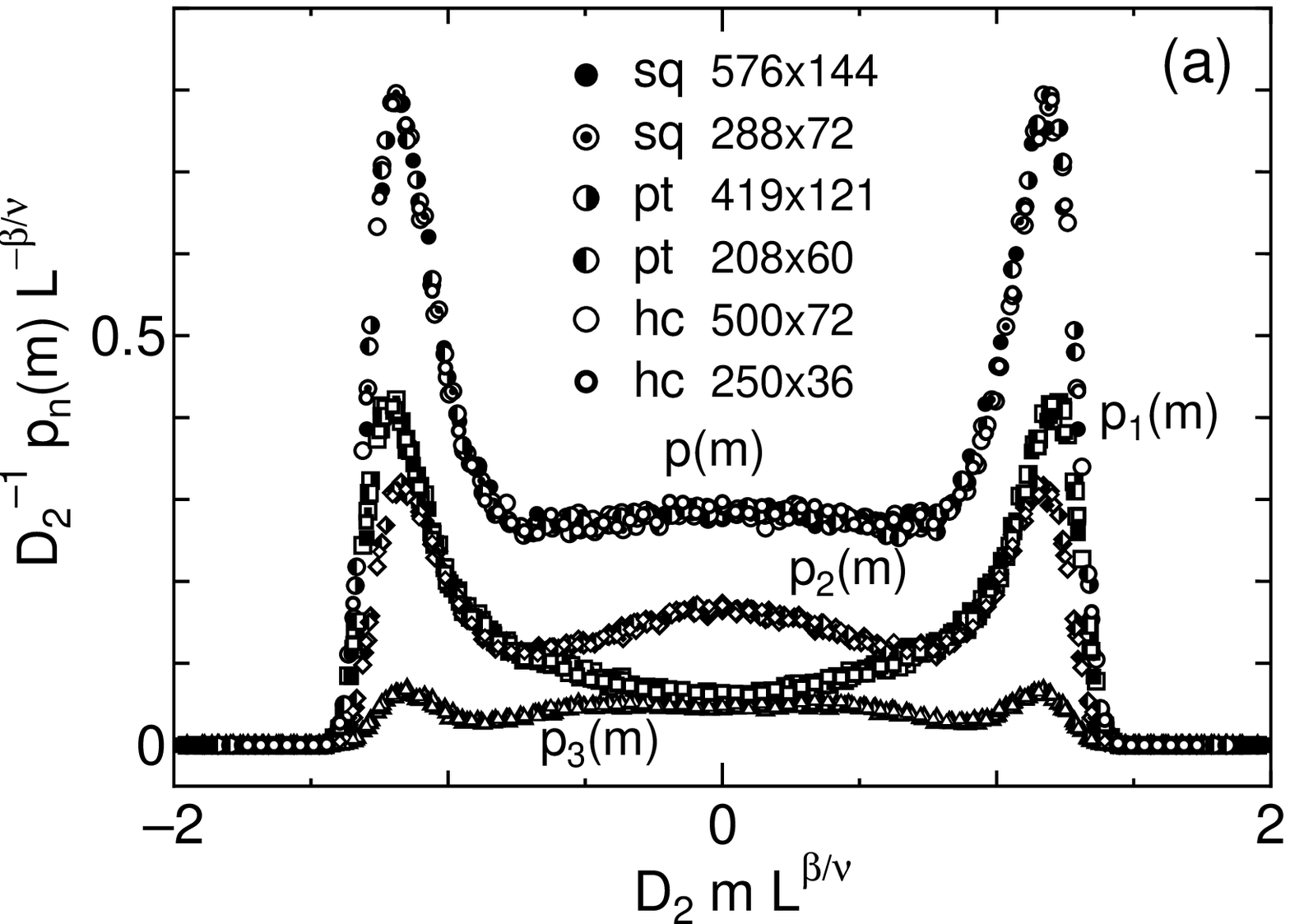}}
\vspace{6mm}
\epsfxsize=\linewidth 
\centerline{\epsfbox{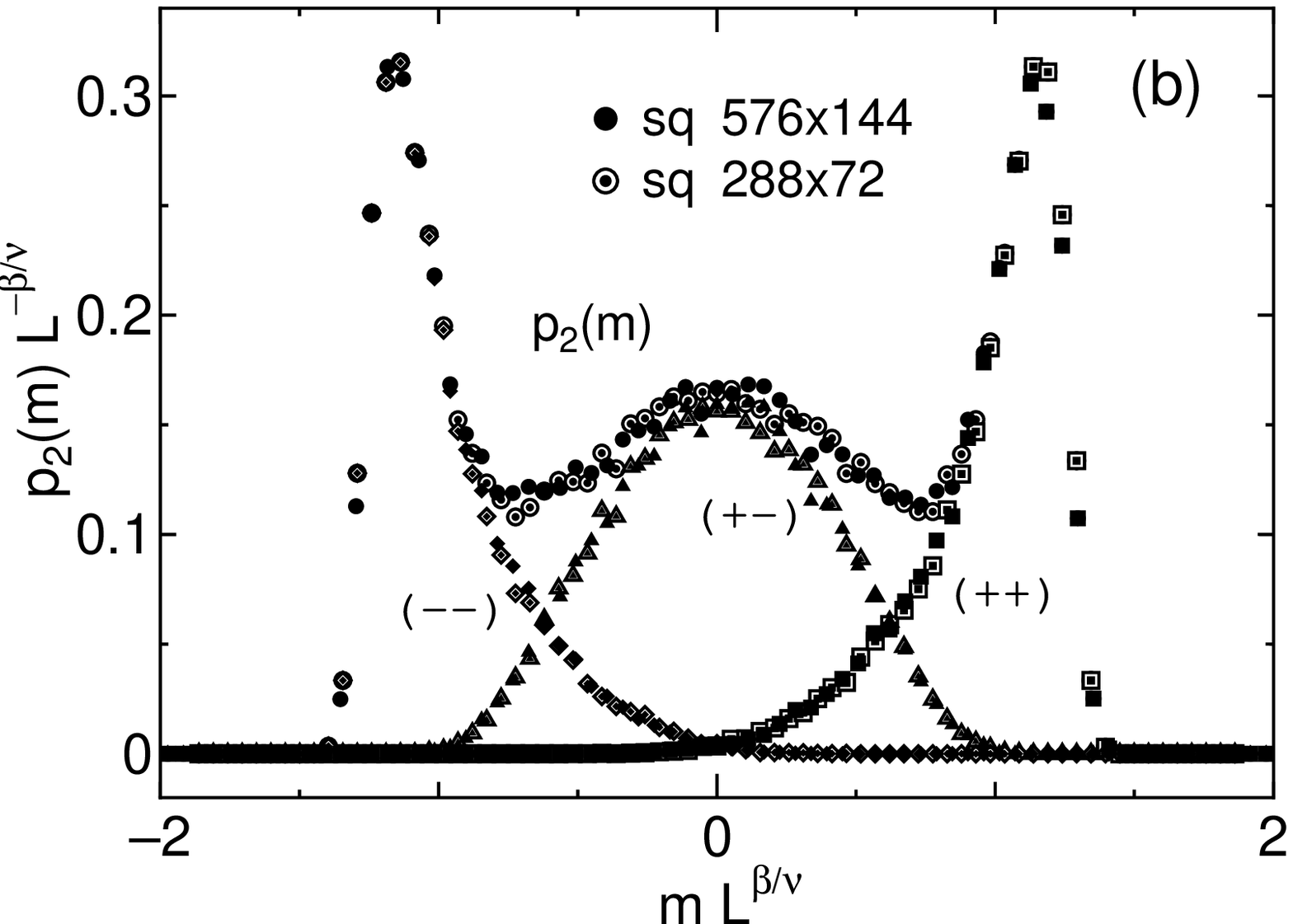}}
\vspace{2mm}
\caption{
(a) $D_2^{-1}p_n(m)L^{-\beta/\nu}$ at $T=T_c$ as a function of
 $D_2 m L^{\beta/\nu}$.
(b) $p_2(m)$ at $T=T_c$ is decomposed into three classes,
$p_{++}(m)$, $p_{--}(m)$ and $p_{+-}(m)$.
}
\label{fig2}
\end{figure}

The calculated $D_2^{-1} f_n(c)L^{-\beta/\nu}$ and 
$D_2^{-1} f(c)L^{-\beta/\nu}$ at the 
critical point as a function of $D_2 c L^{\beta/\nu}$ are shown in 
Fig.~\ref{fig1}(a). 
The lattice sizes are given within the figure.  The aspect ratio is
$a=4$ for sq lattice, and corresponding equivalent ratios for 
other lattices.
The calculated $W_n$, $D_2 \l c \r_n  L^{\beta/\nu}$ 
(also $D_2 \l c \r  L^{\beta/\nu}$), and $g_{cn}$ (also $g_c$) 
as a function of $tL^{1/\nu}$ ($t=(T-T_c)/T_c$)  
are presented in Fig.~\ref{fig1}(b), \ref{fig1}(c), and \ref{fig1}(d), 
respectively.
The calculated $D_2^{-1}p_n(m)L^{-\beta/\nu}$ as a function of 
$D_2 m L^{\beta/\nu}$ is shown in Fig.~\ref{fig2}(a).
The metric factors $D_1$ and $D_2$
for the sq lattice are chosen as 1 \cite{rem_amp}.
The values of $D_1$ for the pt and hc lattices are $1.00 \pm 0.01$ which
are consistent with the results of \cite{hu98};
the values of $D_2$ for the pt and hc lattices are
$1.02 \pm 0.01$ and $0.98 \pm 0.01$, respectively.
Since we have estimated $D_1$ as $1.00 \pm 0.01$ for the
pt and hc lattices, we
have omitted $D_1$ in the horizontal axes of the figures.
Figures \ref{fig1}(a)-\ref{fig1}(d) and Fig.~\ref{fig2}(a) show 
that the calculated quantities 
have very good finite-size scaling behavior and the 
universality is also well satisfied.
We should note that the metric factors $D_1$ and $D_2$
are the same for all quantities.

\begin{figure}
\epsfxsize=\linewidth 
\centerline{\epsfbox{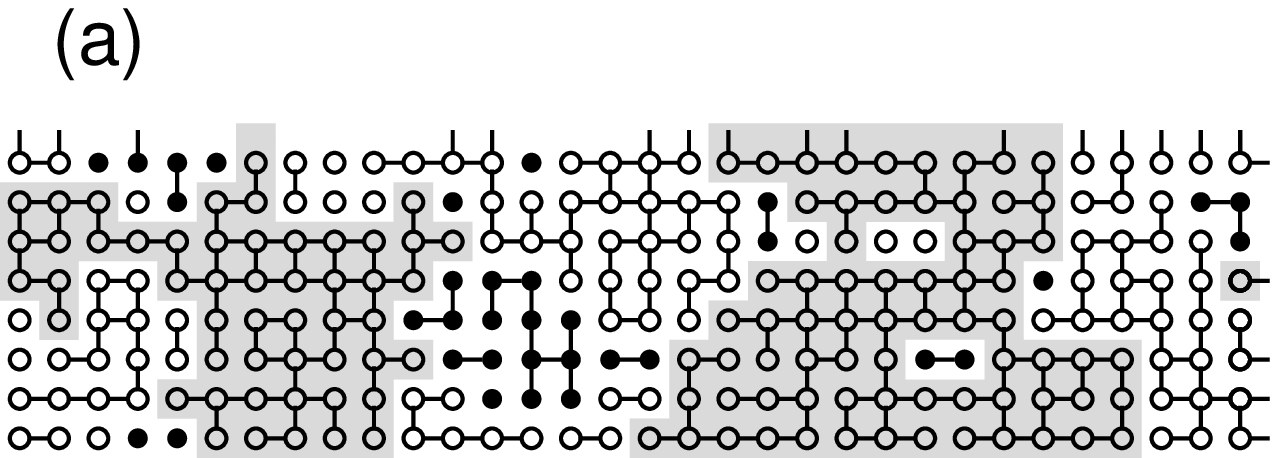}}
\vspace{4mm}
\epsfxsize=\linewidth 
\centerline{\epsfbox{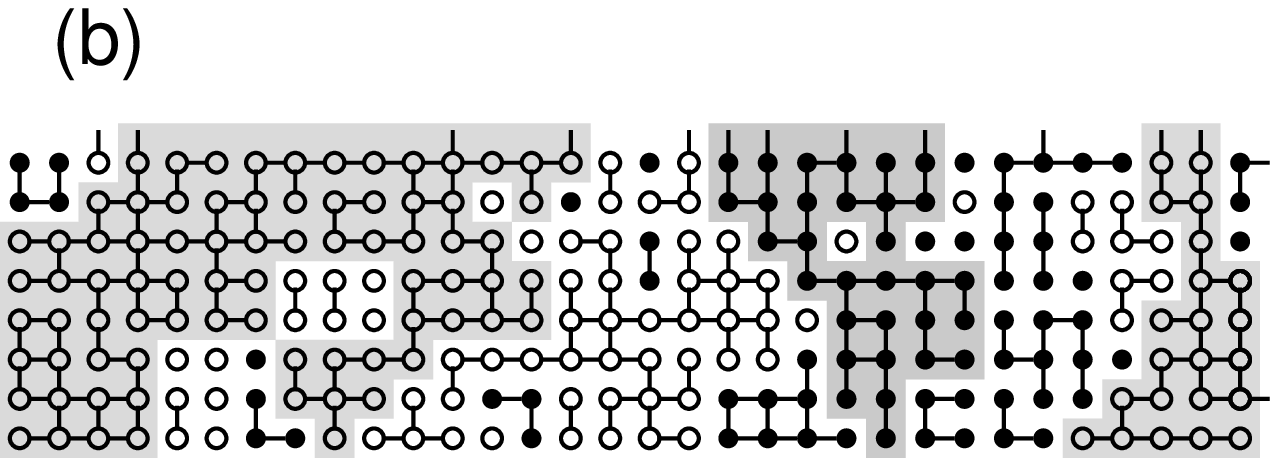}}
\vspace{2mm}
\caption{
Examples of snapshots of the Ising system with the aspect ratio 
$a=4$.  Up and down spins are represented by open and closed circles, 
and active bonds are represented by solid line. 
Percolating clusters are distinguished by shaded area.  
There are two percolating clusters with up spins in (a), whereas 
one percolating cluster is up and the other is down in (b).
}
\label{fig3}
\end{figure}

Figure \ref{fig2}(a) shows that $p(m)$ at $T=T_c$ 
has a broad peak centered at $m=0$ in addition to two peaks of 
positive and negative $m$ for the system with the aspect ratio 
$a=4$ for sq lattice. 
This is contrast to the case of $a=1$ where $p(m)$ 
has only two distinct peaks of positive and negative $m$.  
Such dependence of $p(m)$ on $a$ has already been pointed out 
in Ref.~\cite{ok99}.  From Fig.~\ref{fig2}(a), we see 
that the broad peak of $p(m)$ centered at $m=0$ mainly comes from $p_2(m)$.
There are two types of Ising clusters, that is, the clusters with 
up spins or the clusters with down spins.  Therefore, 
if there are many percolating clusters, the combination of 
the percolating clusters with up spins and those with down spins 
makes it possible that the total magnetization becomes close to 0.  
It is known that the normalized fourth moment of $m$, or 
the Binder parameter, at the critical point depends on the aspect ratio 
\cite{ok99,ka93}.  The origin of such a dependence 
can be attributed to the structure of many percolating clusters.
To clarify this situation, we decompose $p_2(m)$ 
at $T=T_c$ into three classes, $p_{++}(m)$, $p_{--}(m)$ 
and $p_{+-}(m)$ shown in Fig.~\ref{fig2}(b).  
We assign three peaks in $p_2(m)$ by the contribution from 
$p_{++}(m)$, $p_{--}(m)$ and $p_{+-}(m)$. 
Examples of snapshots of the Ising system with two percolating clusters 
are presented in Figs.~\ref{fig3}(a) and \ref{fig3}(b). 
In Fig.~\ref{fig3}(a) both percolating clusters are up; 
in Fig.~\ref{fig3}(b) one percolation cluster is up and another
percolating cluster is down.

\begin{figure}
\epsfxsize=\linewidth 
\centerline{\epsfbox{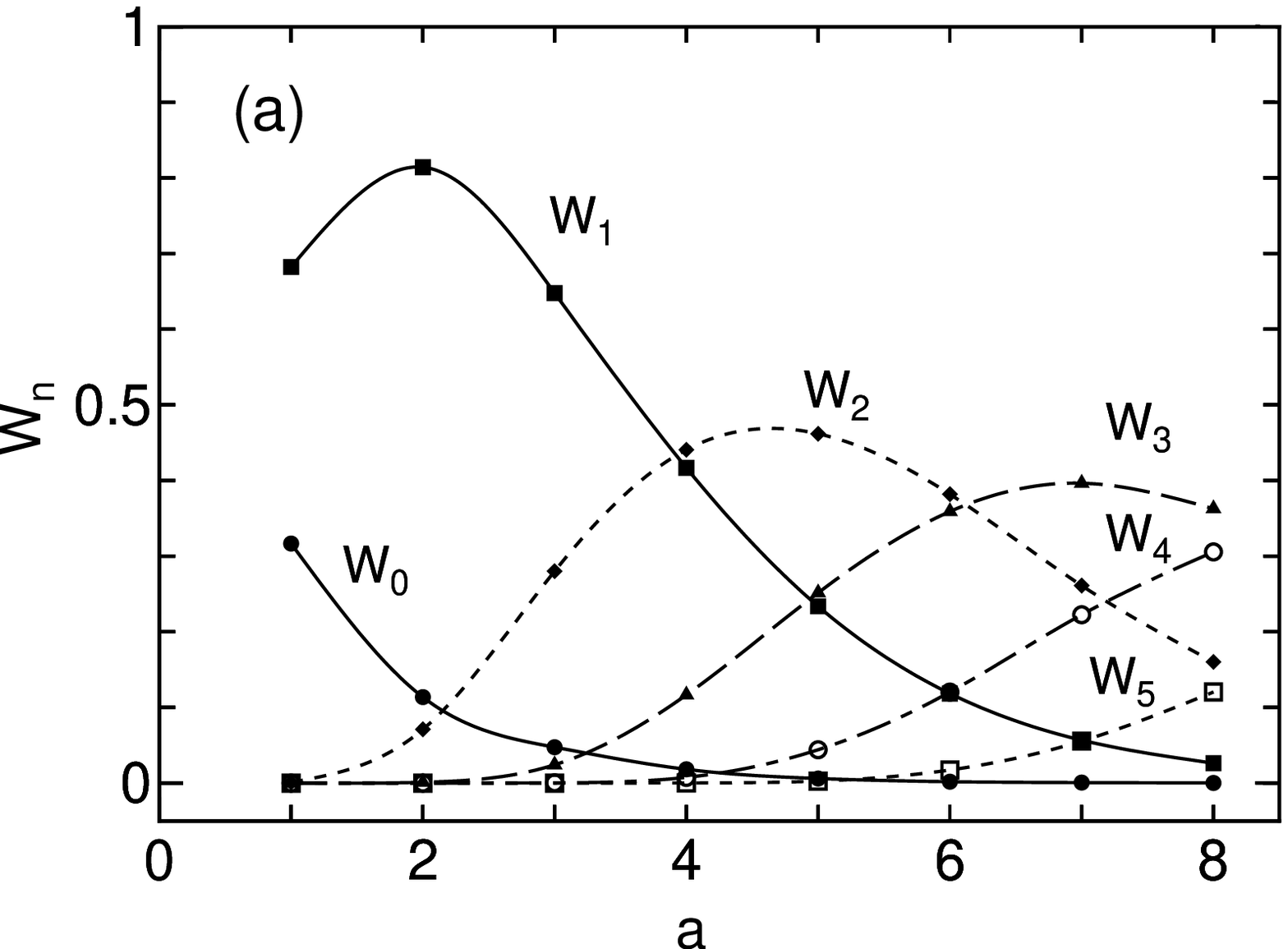}}
\vspace{6mm}
\epsfxsize=\linewidth 
\centerline{\epsfbox{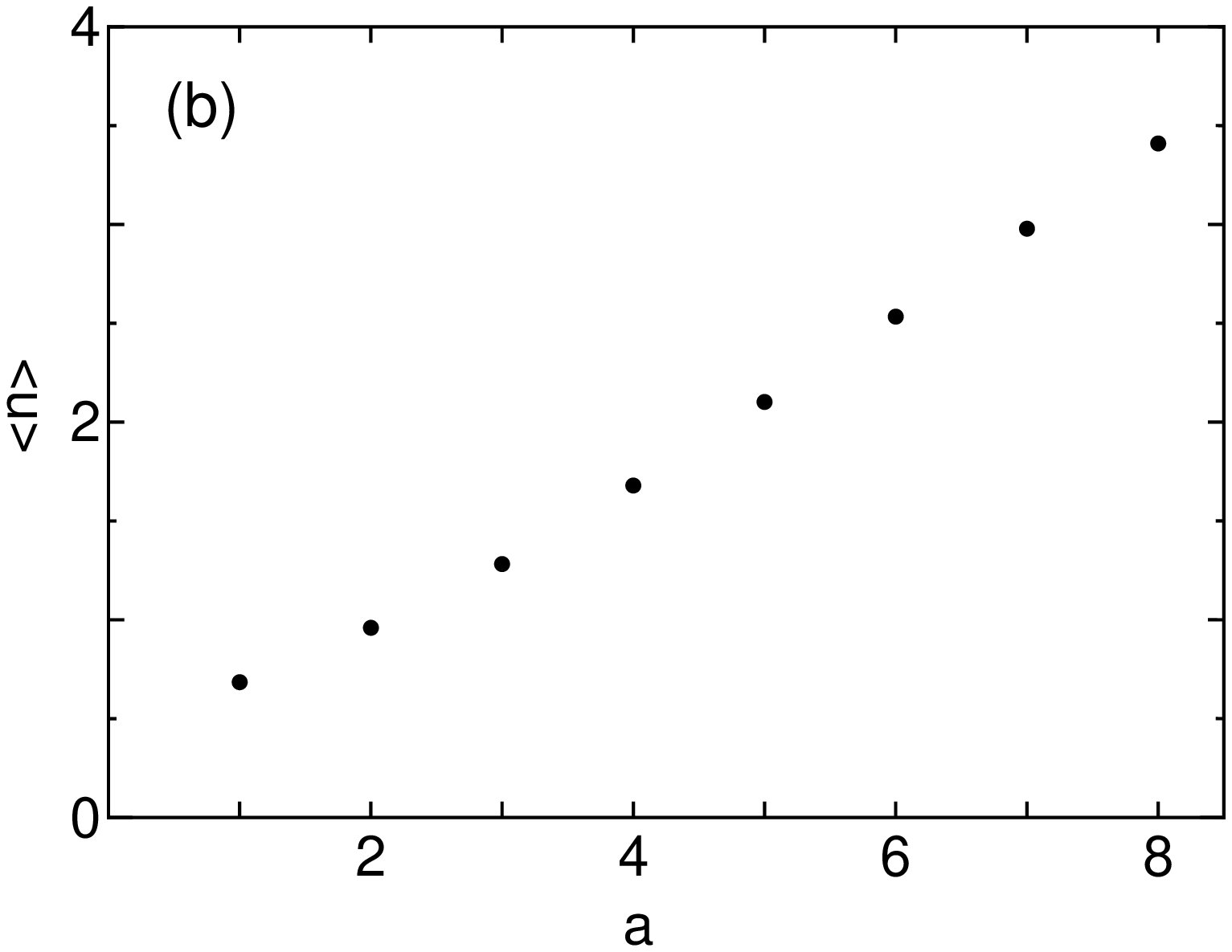}}
\vspace{2mm}
\caption{
(a) $W_n$ at $T=T_c$ as a function of $a = L_1/L_2$.
(b) $\l n \r$ at $T=T_c$ as a function of $a$.
}
\label{fig4}
\end{figure}

From $W_n$, we may calculate the average number of
percolating cluster by $\l n \r=\sum_n nW_n$.  
At the critical point, the values of $W_n$ and $\l n \r$
as a function of the aspect ratio $a=L_1/L_2$ are plotted
in Fig.~\ref{fig4}(a) and \ref{fig4}(b), respectively.
We see that $\l n \r$ increases linearly with $a$ for large $a$, 
which is similar to the case of random percolation \cite{hlc95a,rem_na}. 
The slope of $\l n \r$ versus $a$ in Fig.~\ref{fig4}(b) is approximately 0.5.

Following the study of $W_n$ for random percolation by Hu and Lin
\cite{hlc95a}, there have been many analytic and numerical studies of
$W_n$ in different random percolation problems \cite{mpc}; it is of interest
to extend such studies to the BCPM of the Ising model. On the other hand,
we can extend the study  $f_n(c)$ and $\l c \r_n$ to
the random percolation problem.
It is interesting to compare the results for the BCPM  
and those for the random percolation problem.  
We may also extend the present study to the bond-diluted 
or the site-diluted Ising model which can be mapped
into percolation models \cite{hu91}.
The critical phenomena of the percolating properties of the Ising 
model are governed by the Ising fixed point (for example, 
$\nu = 1$), whereas at the percolation threshold the critical 
phenomena are governed by the random percolation fixed point
($\nu = 4/3$).  The crossover from the Ising fixed point 
to the random percolation fixed point in the process of 
dilution is highly interesting, especially 
for the properties depending on the number of 
the percolating clusters.
The studies in these directions are in progress.

We would like to thank M. Kikuchi and K. Kaneda 
for valuable discussions and the Supercomputer Center
of the ISSP, University of Tokyo, for
providing the computing facilities.
This work was supported by a Grant-in-Aid for Scientific Research 
from the Ministry of Education, Science, Sports and Culture, Japan
and by the National Science Council
of the Republic of China (Taiwan) under grant numbers
NSC 88-2112-M-001-011.


\begin{references}
\bibitem{rem_amp}
 The definition of the metric factor $D_1$ is different
 in Ref.~\cite{ok96}, where $t$ is defined by $T-T_c$.
\end{references}
\end{document}